\documentclass[twocolumn]{aastex62}

\usepackage{graphicx}
\usepackage{bm}
\usepackage{color}      
\usepackage{amsmath}
\usepackage{natbib}
\usepackage{cancel}
\usepackage{soul}

\usepackage{mathrsfs}

\received{2019 July 4}
\revised{--}
\accepted{2019 August 8}
\submitjournal{ApJL}

\shortauthors{L\'{o}pez {\it et al.}}

\begin{document}

\title{Particle-in-cell simulations of the whistler heat-flux instability in 
the solar wind conditions}
\correspondingauthor{R.~A. L\'{o}pez} \email{rlopez186@gmail.com}
\author[0000-0003-3223-1498]{R.~A. L\'{o}pez}
\affiliation{Centre for mathematical Plasma Astrophysics, KU Leuven,
  Celestijnenlaan 200B, B-3001 Leuven, Belgium}
 
\author[0000-0003-0465-598X]{S.~M. Shaaban}
\affiliation{Centre for mathematical Plasma Astrophysics, KU Leuven,
  Celestijnenlaan 200B, B-3001 Leuven, Belgium}
\affiliation{Theoretical Physics Research Group, Physics Department,
  Faculty of Science, Mansoura University, 35516, Mansoura, Egypt}

\author[0000-0002-8508-5466]{M. Lazar}
\affiliation{Centre for mathematical Plasma Astrophysics,
  KU Leuven, Celestijnenlaan 200B, B-3001 Leuven, Belgium}
\affiliation{Institut f\"{u}r Theoretische Physik, Lehrstuhl IV:
  Weltraum- und Astrophysik, Ruhr-Universität Bochum, D-44780 Bochum,
  Germany}

\author[0000-0002-1743-0651]{S. Poedts}
\affiliation{Centre for mathematical Plasma Astrophysics, KU Leuven,
  Celestijnenlaan 200B, B-3001 Leuven, Belgium}

\author[0000-0001-8134-3790]{P.~H. Yoon}
\affiliation{Institute for Physical Science and Technology, University
  of Maryland, College Park, MD, USA}
\affiliation{School of Space Research, Kyung Hee University, Republic
  of Korea}
\affiliation{Korea Astronomy and Space Science Institute, Daejeon
  34055, Republic of Korea}

\author[0000-0001-9293-174X]{A. Micera}
\affiliation{Centre for mathematical Plasma Astrophysics, KU Leuven,
  Celestijnenlaan 200B, B-3001 Leuven, Belgium}
\affiliation{Solar-Terrestrial Centre of Excellence-SIDC, Royal
  Observatory of Belgium, 1180 Brussels, Belgium}

\author[0000-0002-3123-4024]{G. Lapenta}
\affiliation{Centre for mathematical Plasma Astrophysics, KU Leuven,
  Celestijnenlaan 200B, B-3001 Leuven, Belgium}

\begin{abstract}
In collision-poor plasmas from space, e.g., solar wind or stellar
outflows, the heat-flux carried by the strahl or beaming electrons is
expected to be regulated by the self-generated instabilities.
Recently, simultaneous field and particle observations have indeed
revealed enhanced whistler-like fluctuations in the presence of
counter-beaming populations of electrons, connecting these
fluctuations to the whistler heat-flux instability (WHFI). This
instability is predicted only for limited conditions of electron
beam-plasmas, and was not captured in numerical simulations yet. In
this letter we report the first simulations of WHFI in
particle-in-cell (PIC) setups, realistic for the solar wind
conditions, and without temperature gradients or anisotropies to
trigger the instability in the initiation phase. The velocity
distributions have a complex reaction to the enhanced whistler
fluctuations conditioning the instability saturation by a decrease of
the relative drifts combined with induced (effective) temperature
anisotropies (heating the core electrons and pitch-angle and energy
scattering the strahl). These results are in good agreement with a
recent quasilinear approach, and support therefore a largely accepted
belief that WHFI saturates at moderate amplitudes. In anti-sunward
direction the strahl becomes skewed with a pitch-angle distribution
decreasing in width as electron energy increases, that seems to be
characteristic to self-generated whistlers and not to small-scale
turbulence.

\end{abstract}


\section{Motivations}

Among the kinetic instabilities invoked in the self-regulation of
solar wind properties the heat-flux instabilities, and in particular
the whistler heat-flux instability (WHFI), are still the most
controversial, though in the last decade an increased effort has been
devoted to understanding their fundamental properties
\citep{Saito2007a, Pavan2013, Seough2015a, Saeed2017a, Saeed2017b,
  Shaaban2018a, Shaaban2018b, Shaaban2019, Lee2019} and find their
signatures in the observation \citep{Breneman2010, Gurgiolo2012,
  Wilson2013, Landi2014, Lacombe2014, Stansby2016, Tong2019a,
  Tong2019b}. The WHFI is triggered by the relative drift, $U= |U_c| +
U_b$, of the counter-beaming electrons, a central population (summing
up the core and halo electrons) here called generically core and
denoted by subscript $c$, and the beam or strahl population (with
subscript $b$) satisfying the zero net-current condition $n_c |U_c| =
n_b U_b$, see \citet{Gary1985} and refs therein. Conditions for the
whistlers to be excited (resonantly) by the beaming electrons are
however very restrained, namely, to a beaming velocity limited between
two threshold values, roughly given by $\theta_c < U_b < \theta_b$,
where $\theta_{c,b}$ are thermal velocities \citep{Gary1985,
  Shaaban2018a, Shaaban2018b}. The quasi-stable states are expected in
this case only for low drifts $U_b$ (or $U_c$), below the lower
threshold \citep{Gary1999a, Shaaban2018a} that seems to be confirmed
by the observations \citep{Gary1999a, Gary1999b,
  Tong2018}. Theoretically, whistlers may also satisfy resonance
conditions with both electron populations, especially, for more
energetic beams ($U_b > \theta_b$), but never develop, being heavily
competed by the other faster growing modes, e.g., the electrostatic
beam-plasma instabilities or the oblique instabilities
\citep{Gary2007, Saito2007a, Seough2015a, Saeed2017a, Horaites2018,
  Lee2019, Vasko2019, Verscharen2019b}. If the core electrons exhibit
an important temperature anisotropy $T_{c,\perp} > T_{c,\parallel}$
the regime of WHFI may be significantly altered becoming specific to a
standard whistler instability driven by temperature anisotropy, with
lower thresholds and higher growth rates \citep{Seough2015a,
  Shaaban2018b}.

The first investigations of WHFI have been stimulated by the
observations suggesting a potential implication of whistlers in the
regulation of suprathermal populations. If binary collisions are rare,
in the solar wind the electron heat-flux is less than a conventional
Spitzer-H{\"a}rm level \citep{Spitzer1953}, and such constraint is
attributed mainly to the wave-particle interactions \citep{Bale2013}.
Moreover, with the expansion of the solar wind the electron halo shows
a continuous build-up on the expense of strahl that lowers in
intensity and undergoes a pitch-angle scattering
\citep{Maksimovic2005, Pagel2007, Gurgiolo2012, Bercic2019}. In the
absence of collisions an immediate explanation for these evolutions is
offered by the small scale wave turbulence and/or the fluctuations
self-generated by the instabilities. Higher plasma beta conditions
stimulate the implication of self-generated instabilities in the
regulation of suprathermal populations, in particular of the electron
strahls \citep{Pilipp1987, Crooker2003}.

Theory and simulations have confirmed that whistler fluctuations,
either predefined by a power spectrum decreasing monotonically with
increasing frequency or self-generated by kinetic instabilities, can
pitch-angle and energy scatter the suprathermal electrons and lead to
asymmetric beaming-like distributions, broader (bulge) or decreasing
(skewness) in pitch-angle at larger electron energies
\citep{Vocks2003, Vocks2005, Saito2007a, Saito2007b, Seough2015a}.  In
particular, for the WHFI, quasilinear studies have also suggested a
potential role in the limitation of the electron heat-flux, probably
by the same mechanisms, which reduce the relative drift and induce
effective anisotropies of electron populations \citep{Gary1977,
  Shaaban2019}. However, a confirmation of these effects in
simulations have not been reported yet. To our knowledge, numerical
experiments have provided extended descriptions only for other
different branches of heat-flux instabilities, e.g., electrostatic
beam-plasma, firehose-like \citep{Gary2007, Lee2019}, or for the
temperature anisotropy driven instabilities \citep{Saito2007a,
  Seough2015a}.

Limiting conditions predicted for the WHFI \citep{Gary1985,
  Shaaban2018a} and small amplitudes of the resulting fluctuations
\citep{Shaaban2019, Tong2019b} might have also prevented a direct
detection in the observations, and leaded sometimes to contradictory
correlations between plasma states and fluctuations
\citep{Scime2001}. Clear evidences of WHFI in the solar wind have
recently been provided by simultaneous electron and field measurements
with a well established connection to the electron counter-beaming
populations and their temperature anisotropy \citep{Tong2019a,
  Tong2019b}. These observations confirm recent predictions that WHFI
must be quenched by a slight anisotropy $T_{b,\parallel} \gtrsim
T_{b,\perp}$ of the beam, but growth rates may significantly be
increased by an opposite anisotropy of the core $T_{c,\perp} \gtrsim
T_{c,\parallel}$ \citep{Shaaban2018b}.

This letter reports the first particle-in-cell (PIC) simulations of
the WHFI, realistic for solar wind conditions. The characteristics of
this instability (see above) impose serious limitations to describe it
using simulations (with realistic parameters), requiring an immense
amount of numerical resources, which are practically impossibles with
standard approaches. Here we make use of an implicit PIC code
developed by \citet{Markidis2010}, able to resolve multiple temporal
and spatial scales characteristic to the solar wind plasma dynamics
\citep{Verscharen2019}, without the strict limitations in time step
and grid spacing imposed typically in explicit codes. Simulations
capture the energy transfer between the electron core and beam
populations, and correctly describe the saturation of WHFI via the
relaxation of the velocity distributions.  A quasilinear approach
allows time variations of the moments of the distribution (e.g.,
drifts, temperatures), but implies only a single wave mode in the
energy and momentum transfers \citep{Shaaban2019}. Instead the
simulations enable quasilinear and nonlinear effects of multiple
(concurrent, coupled) wave modes.

\section{Particle-in-cell simulations} \label{sec:results}

\begin{figure*}[h!t]
  \begin{center}
    \includegraphics[width=0.4\textwidth]{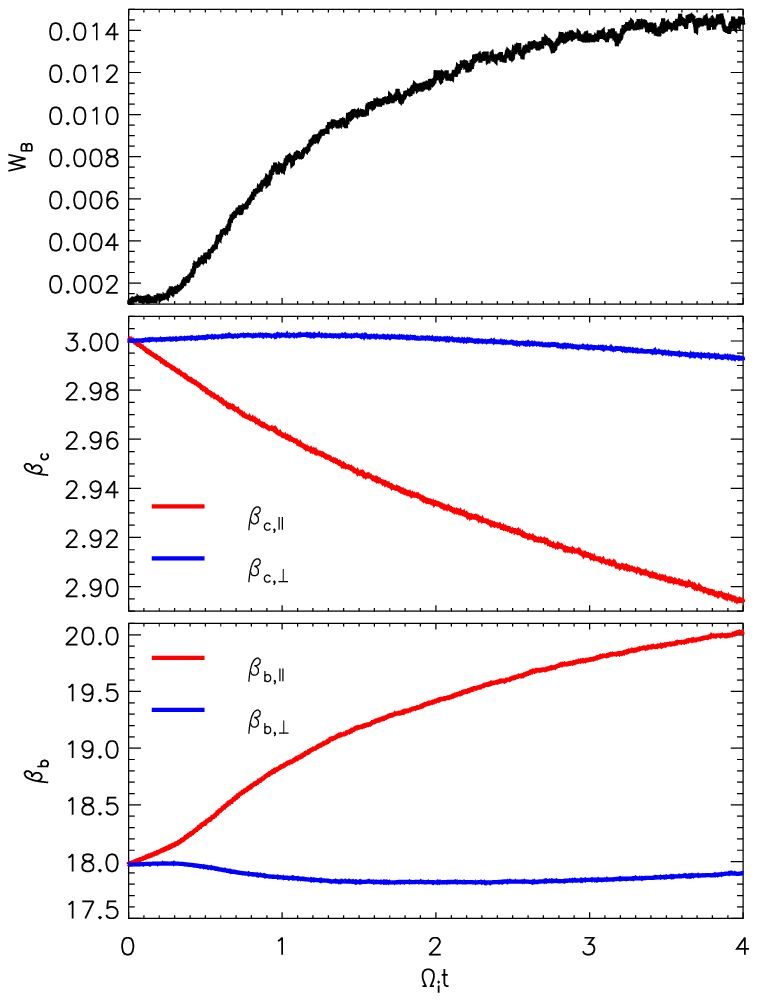}
    \includegraphics[width=0.4\textwidth]{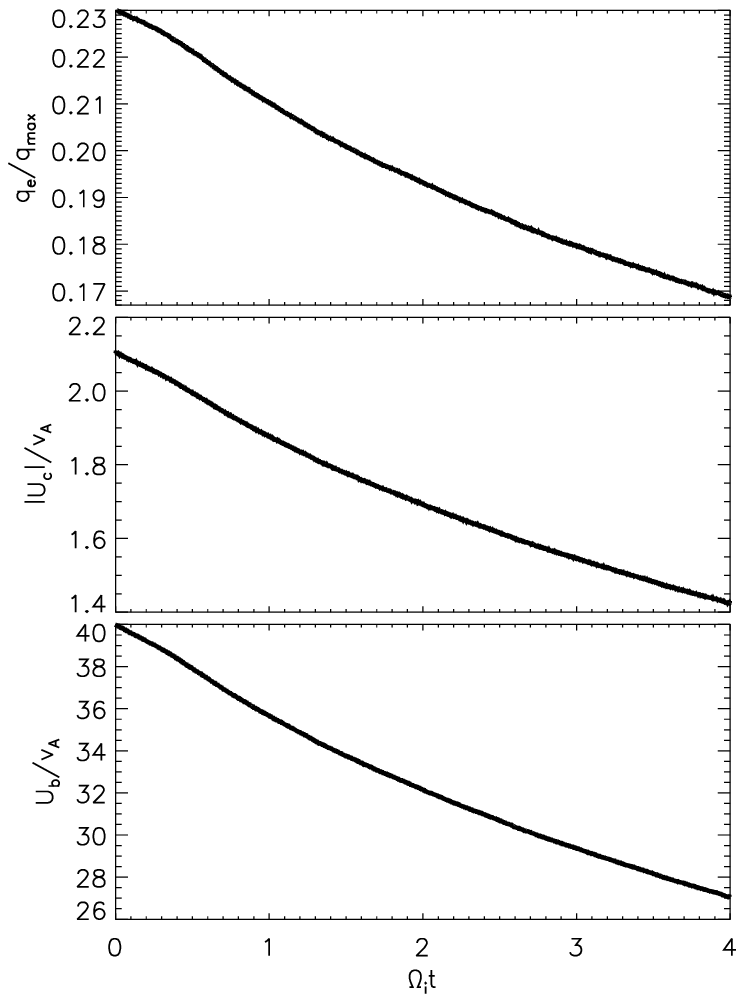}
    \caption{Temporal evolution for the fluctuating magnetic energy
      density $W_B$, parallel and perpendicular
      components of plasma beta parameters $\beta_{c,b} $, normalized
      (parallel) electron heat-flux, and parallel drifts
      $U_{c,b}$.}
    \label{fig:moments}
  \end{center}
\end{figure*} 
\begin{figure}[h!t]
  \begin{center} 
    \includegraphics[width=0.45\textwidth]{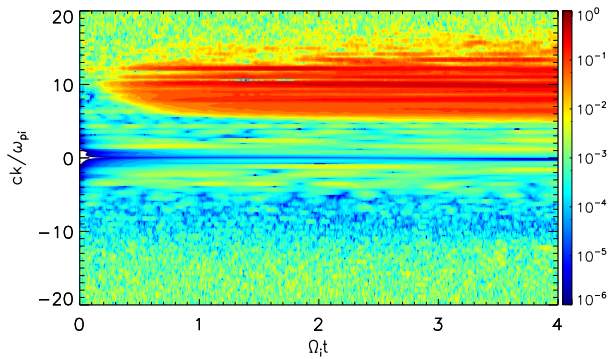} 
    \caption{Temporal evolution of the wave number transverse magnetic
      power.} \label{fig:kt}
  \end{center}
\end{figure}

Our initial setup in Table \ref{t1} is intended to the solar wind
plasma conditions \citep{Maksimovic2005, Tong2018, Tong2019a}, and,
for simplicity, both counter-beaming electron populations (in a frame
fixed to protons)
\begin{align}
    f_e\left({ v_{\perp}, v_{\parallel}}\right) =
    \frac{n_c}{n_e}~f_c\left({ v_{\perp}, v_{\parallel}}\right)
    +\frac{n_b}{n_e}~f_b\left({ v_{\perp}, v_{\parallel}}\right)
\end{align}
are assumed Maxwellian distributed. Here $n_c$ and $n_b$ are the core
and beam number densities, respectively, and $n_e\equiv n_0$ is the
total number density of electrons, in a neutral plasma with zero
charge $n_e \approx n_i$ and zero net current $n_c U_c + n_b U_b = 0$,
where $U_{b,c}$ are the corresponding drift velocities, and here the
ions (subscript $i$) are assumed to be only protons.

\setlength{\tabcolsep}{8pt}
\begin{deluxetable}{lccc}
    \tablenum{1}
    \tablecaption{Initial plasma parameters for the simulation.\label{t1}}
    \tablehead{Parameter & Beam ($b$)  & Core ($c$) & Protons ($p$)}
    \startdata
        $n_j/n_0$  &0.05 & 0.95 & 1.0 \\
        $T_{j,\parallel}/T_{c,\parallel}$ & 6.0 & 1.0 & 1.0 \\
        $\beta_{j,\parallel}$ & 18.0 & 3.0 & 3.0 \\
        $m_p/m_j$ & 1836 & 1836 & 1.0\\
        $T_{j,\perp}/T_{j,\parallel}$ & 1.0 & 1.0 & 1.0\\
        $U_j/v_A$ &  $40.0$ & $-2.1$ & 0.0
    \enddata
\end{deluxetable}

We use an implicit one-dimensional PIC code, i.e.,
iPic3D~\citep{Markidis2010} with a high enough resolution to resolve
the electron inertial length and the electron gyromotion.  The spatial
grid is composed of $n_x= 1024$ cells, with $5000$ particles per
species per grid. The box size is $L_x=16\,d_i$, then the cell size is
$\Delta x=0.0156\,d_i$. Here $d_i=c/\omega_{pi}$ is the ion inertial
length, with $\omega_{pi}=(4\pi n_0e^2/m_p)^{1/2}$ the ion plasma
frequency.  The mass ratio is $m_p/m_e=1836$, and the plasma to gyro
frequency ratio for ions is $\omega_{pi}/\Omega_{ci}=4390.07$, which
implies that the Alfv\'en speed is $v_A = B_0/\sqrt{4\pi n_pm_p} =
0.00023\,c$ and the plasma to gyro frequency of the electrons,
$\omega_{pe}/\Omega_{ce}=102.48$, which are typical values encounter
under solar wind conditions. The background magnetic field is set in
the $x$ direction, $\mathbf{B}_0 = B_0\hat{\mathbf{x}}$. The time step
is $\Delta t=0.0375/\omega_{pi}$ and the simulations ran until
$t_\text{max}=17560.265/\omega_{pi}$ or equivalently
$t_\text{max}=4.0/\Omega_{ci}$. In terms of electron quantities, the
time step used correspond to $\Delta t=0.016/\Omega_{ce}$ and the cell
size is $\Delta x=0.7\,d_e$. 

\begin{figure}[h!t]
  \begin{center} 
    \includegraphics[width=0.45\textwidth, trim={0 0.2cm 0 0}, clip]{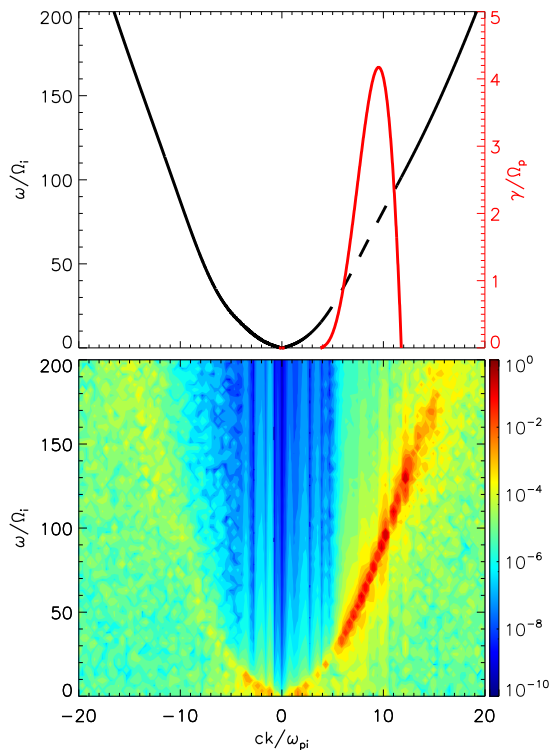} 
    \caption{Qualitative comparison of linear dispersion relation
      (upper panel), real frequency (black, dashed line representing
      the unstable region) and growth rate (red), with normalized
      power spectra of whistler fluctuations for the interval
      $0<\omega_it<2.0$.} \label{fig:disp}
  \end{center}
\end{figure}

Fig.~\ref{fig:moments} shows (normalized) time variations, with the
increase and saturation of the magnetic power ($W_B=\int\delta
B^2/B_0^2\,dx$) of the enhanced fluctuations, and the relaxation of
the main moments of electron velocity distributions which continues
after the instability saturation. The rapid growth of $W_B$
corresponding to the excitation of WHFI in the early stage of the
simulations slows down close to $\Omega_it \approx 3$, and shows then
a slower increase up to the end of the simulation, $\Omega_p t =
4.0$. The entire period of the simulation can be identified in this
case as characteristic to a pure WHFI. In order to identify this
interval of pure whistler-like fluctuations we have used the fast
Fourier transforms in space of the transverse magnetic fluctuations
($|\text{FFT}(B_y-iB_z)|^2$), which are displayed in
Fig.~\ref{fig:kt}. In this interval only the intense power of the WHFI
corresponding to positive wave numbers are present, see also
Fig.~\ref{fig:disp}.
\begin{figure*}[t]
  \begin{center}
    \includegraphics[width=0.9\textwidth]{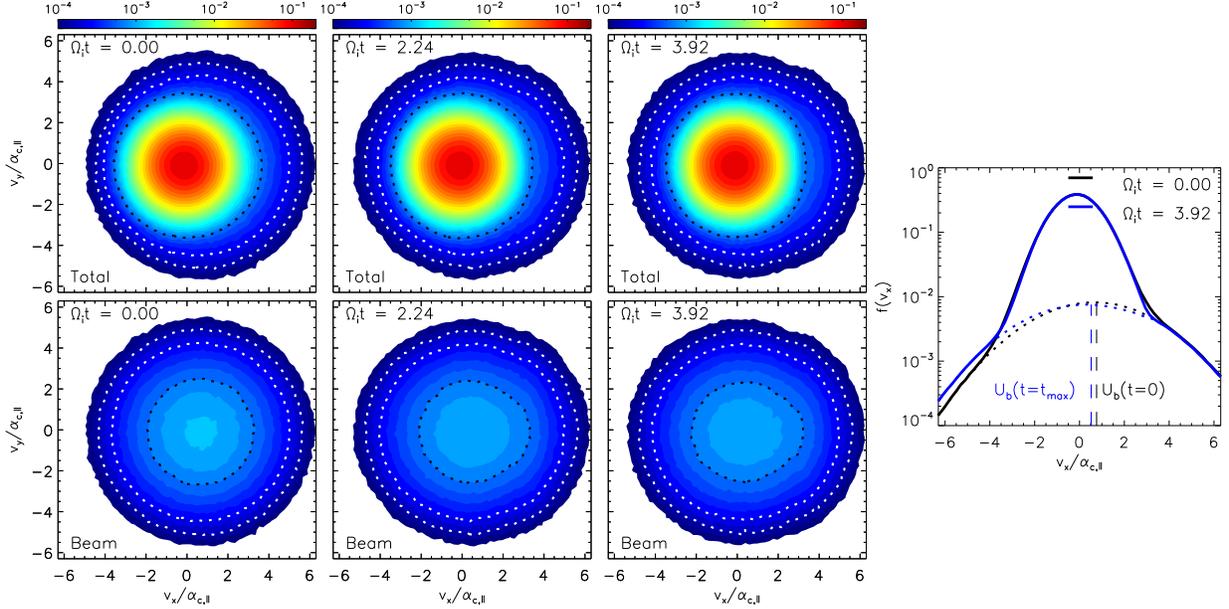}
    \caption{The eVDF $f_e(v_x,v_y)$ at different stages in the
      simulation, $\Omega_it=0.0$ $3.3$ and $10.1$. Upper panels are
      showing the total eVDF and lower panels only the beam
      distributions.  Initial ($\Omega_it=0.0$; black) and final
      ($\Omega_it=10.1$; blue) snapshots of the reduced eVDF
      $f_e(v_x)$ (right panel).}\label{fig:fhist}
  \end{center}
\end{figure*}

Plasma beta parameters $\beta_{c,b}\equiv 8\pi n_0 T_{c,b}/B_0^2$
plotted in Fig.~\ref{fig:moments} are defined with total number
density $n_0$ and reflect therefore the variations of the
corresponding temperatures $T_{c,b}$, in parallel (red) and
perpendicular (blue) directions with respect to the background
magnetic field. Initially isotropic, i.e., $\beta_{c,\parallel}(0)=
\beta_{c,\perp}(0)$, the core temperatures are subjected to parallel
cooling (red) and small perpendicular heating (blue) by a resonant
cyclotron interaction of whistlers with the cooler electrons from the
core. Beaming electrons have also isotropic temperatures at the
beginning, i.e., $\beta_{b,\parallel}(0)=\beta_{b,\perp}(0)$, but
their pitch-angle and energy scattering induces an opposite
anisotropy. Consequently, at the saturation the core exhibits an
excess of perpendicular temperature, i.e. $\beta_{c,\perp}(t^*) >
\beta_{c, \parallel}(t^* )$, while the beam shows an excess of
parallel temperature, i.e.\ $\beta_{b,\perp}(t^*) < \beta_{b,
  \parallel}(t^*)$. Both the linear theory of WHFI
\citep{Shaaban2018b} and the observations \citep{Tong2019a} suggest
indeed that this instability is inhibited by such a temperature
anisotropy of the beam. These results are also in good agreement, at
least at qualitative level and for the same time scales, with the QL
evolutions predicted by theory, see, for instance, Fig.~9 (middle
panels) and Fig.~10 in \citet{Shaaban2019}. However, for such
comparison we have to keep in mind that initial conditions, like the
number of particles used in PIC simulations, or the initial level of
the electromagnetic fluctuations in the QL theory, or both are crucial
for the onset time of the instability \citep{Lopez2018}.
  
Right panels in Fig.~\ref{fig:moments} show the time relaxation of the
electron heat-flux $q_e =m_e/2\int dv\,v_x v f_e$ (normalized by
$q_\text{max}=3 n_0 T_{\parallel c}\alpha_{c,\parallel}/\sqrt{2}$,
where $\alpha_{c,\parallel}=\sqrt{k_B T_{\parallel c}/m_e}$ is the
thermal speed), and the core and beam (normalized) drift velocities
($U_{c,b}/v_A$). In the time interval relevant for the WHFI the heat
flux and (counter-)drifts are only partially relaxed, showing similar
reductions of about $25\%$ or $30\%$ of initial magnitude. We can
state that the relaxation of relative drift velocities,
i.e.\ $U_{c,b}(t_\text{max})\approx~0.67~U_{c,b}(0)$, is slowed down
by a concurrent effect of the enhanced fluctuations, which interact
with the electrons and induce opposite temperature anisotropies in the
core and beam populations.

In Fig.~\ref{fig:disp} we plot the normalized power spectra for the
initial stage of the simulation, i.e.\ $0<\Omega_it<2.24$, to
guarantee we are capturing the linear stage of the WHF instability and
to have a fair comparison with the linear dispersion relation at
$\Omega_it=0$. The spectra is obtained from
$|\text{FFT}(B_y-iB_z)|^2$, where here the FFT is computed in space
and time (then normalized to the maximum value of the spectra). By
doing so, we are able to separate the contribution of left-handed (LH)
and RH circularly polarized modes. Thus, for $\omega>0$ we observe the
RH contribution with positive and negative helicity, $k>0$ and $k<0$,
respectively, see \citet{Saeed2017a} for details. Most of the magnetic
power is concentrated in the part with $\omega>0$ and $k>0$,
corresponding to the RH unstable modes with positive helicity,
confirming the linear theory predictions (top panel) for a RH
WHFI. Moreover, we observe a very low intensity in the negative
wave-numbers part of the spectrum, but those are modes with negative
helicity and are damped, according to linear calculations. The other
combinations do not show any significant power (not shown here). Here
we can state that our PIC simulations are capable to capture the low
intensity whistler fluctuations associated with the WHFI that can
develop only for $\omega>0$ and $k>0$. The dispersion shown by the
simulated fluctuations is not an instantaneous picture, but rather a
cumulative contribution of fluctuations in the entire period under
consideration, when macroscopic plasma values evolve from the initial
condition.  However, the unstable wavenumber interval does not change
much with the beaming speed (see, Figs. 1 and 2 in
\citet{Shaaban2019}) to explain the broad wavenumber spectra, which
may probably result from the small error in the energy conservation in
this simulation. Reducing the time step or increasing the number of
particles per grid cell would help to improve the energy conservation
and therefore obtain more accurate results, but more computational
resources will be needed.

\begin{figure}[h!]
  \begin{center}
    \includegraphics[width=0.45\textwidth]{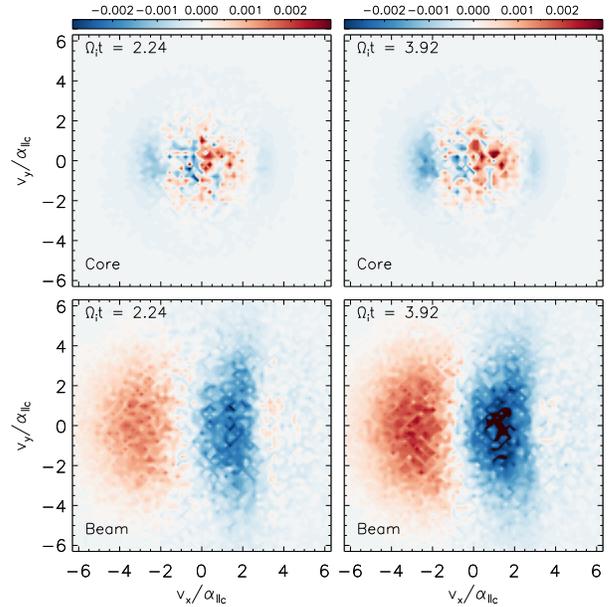}
    \caption{Fluctuating distribution function $\delta f_j(t) =
      f_j(t)-f_j(0)$: core distribution $\delta f_c$ (top) and beam
      distribution $\delta f_b$ (bottom).}
    \label{fig:df}
  \end{center}
\end{figure}

Fig.~\ref{fig:fhist} presents the velocity distribution $f(v_x,v_y)$
at different relevant stages of the simulation
$\Omega_it=0.0$, $2.24$, and $3.92$, for the total
electron population (upper panels) and the beam component (lower
panels), as well as the reduced distribution (integrated along $v_y$)
$f_e(v_x)$ at the initial and (almost) final stages of the simulation,
i.e.\ $\Omega_it=0.0$, and $3.92$ (right panel). In order to highlight
the deformation of the electron components in the distribution, we
have carefully selected three particular contours, as indicated with
dotted lines, at $2 \times 10^{-4},\,3\times 10^{-4},\,8\times
10^{-4}$. It is clear that the highest contour of level $8 \times
10^{-4}$ (black dotted) becomes more symmetric at the end of the
simulation, showing also a slight increase of temperature anisotropy
of the core population $T_{c,\perp} > T_{c,\parallel}$ (upper panels)
and giving an indication for the relaxation of the drift
velocities. In the case of the beam (lower panels) this contour shows
the behavior observed in Fig.~\ref{fig:moments}, a generation of
parallel anisotropy.  At later stages of the simulation contours of
lower level (white), e.g., $3\times 10^{-4}$ and $\,2 \times 10^{-4}$,
are slightly different than those at the initial state, and
specifically show an asymmetric skewness of less scattered particles
(pitch-angle scattering of the beam decreasing in parallel direction
as electron energy increases). Moreover, a lower (relaxed) but still
finite drift velocity is more obviously shown by the reduced
distributions in the right panel. The reduced distributions confirm
the previous description that the initial drift velocities (black
line) are regulated by the enhanced WHF fluctuations and the electron
components ended up with small but finite relative drift velocities
(blue line). Moreover, at final stage, i.e.\ $\Omega_it=3.92$, the
reduced eVDF $f_e(v_x)$ shows the formation of a small, but still
noticeable, ``shoulder'' in the parallel direction for the beam
component, suggesting already that not all beaming electrons are
scattered by the enhanced fluctuations, an hypothesis confirmed by the
results in Fig.~\ref{fig:df}.

Finally, Fig.~\ref{fig:df} shows the departures of the distributions
from the initial condition $\delta f_j(t)=f_j(t)-f_j(0)$, for core
(top panels) and beam (bottom panels) electrons, and for the two
relevant moments $\Omega_it=2.24$ (left) and $\Omega_it=3.92$
(right). Red contours show $\delta f_j>0$ with an abundance of
scattered electrons, while blue contours $\delta f_j<0$ mark the
electron loss. Here we can see how different electron components are
scattered (or not) by the whistler waves. Correlating with
Fig.~\ref{fig:moments}, the diffusion of core electrons occurs
under the effect of whistlers which interact
resonantly with the electrons with $v_x <|U_c|$, while the instability
itself is (resonantly) triggered by the beaming electrons with $v_x <
U_b$, cooling them down in perpendicular direction and increasing
their effective temperature (or kinetic energy) in parallel
direction.
The lighter blue color population at higher energies in
Fig.~\ref{fig:df}, right-lower panel, indicates those electrons less
scattered by whistlers, and corresponds to the small shoulder (or
small plateau) showed in Fig.~\ref{fig:fhist}. In time this population
is naturally reduced leading to a lower pitch-angular width that
becomes however prominent due to a concomitant decrease of the drift.

%

\section{Summary} \label{sec:summary}

In this Letter we have provided a detailed description of the whistler
heat-flux instability (WHFI) using an implicit one-dimensional PIC
simulation. The instability is triggered by the relative drift of the
counter-beaming electron populations, without temperature gradients or
temperature anisotropies. The initial stage of the simulation is
characterized by a rapid growth of the magnetic energy density,
corresponding to the excitation of the WHFI, then corroborated by the
spectral analysis with a good agreement with linear theory. The
enhanced whistler fluctuations interact with both electron components,
reducing the relative drift ($\sim30 \%$) and inducing (effective)
temperature anisotropies, i.e., an excess of perpendicular temperature
for the core and excess of parallel temperature for the beam.
The interplay of temperature anisotropies is in good agreement with a
recent QL approach \citep{Shaaban2019} (although the drift relaxation
is less significant under the effect of a single mode in QL theory),
and support therefore a largely accepted belief that WHFI saturates at
typically small amplitudes.

In anti-sunward direction the strahl becomes skewed with a pitch-angle
distribution decreasing in width as electron energy increases, that
seems to be characteristic to self-generated whistlers and not to
small-scale turbulence.  However, this skewness (a decreasing
pitch-angle distribution with increasing energy) is shown only by the
lower levels (white dashed contours) with a lower contribution to the
moments of the distribution, and implicitly to the effective
temperature anisotropy, which is reduced.

Future refinements to clarify the nonlinear evolution of this
instability need to be taken with caution and eventually using new
codes that conserve much better the energy.

\acknowledgments These results were obtained in the framework of the
projects SCHL~201/35-1 (DFG-German Research Foundation), GOA/2015-014
(KU Leuven), G0A2316N (FWO-Vlaanderen), and C~90347 (ESA Prodex
9). S.M.S. acknowledges support by a FWO Postdoctoral Fellowship,
Grant No.~12Z6218N. P.H.Y. acknowledge BK21 Plus program from NRF
Korea. The computational resources and services used in this work were
provided by the VSC (Flemish Supercomputer Center), funded by the
Research Foundation - Flanders (FWO) and the Flemish Government --
department EWI. We acknowledge fruitful discussions at the meeting of
international team on Kappa Distributions hosted by ISSI-Bern.
 
  

\begin{thebibliography}{}
\expandafter\ifx\csname natexlab\endcsname\relax\def\natexlab#1{#1}\fi
\providecommand{\url}[1]{\href{#1}{#1}}
\providecommand{\dodoi}[1]{doi:~\href{http://doi.org/#1}{\nolinkurl{#1}}}
\providecommand{\doeprint}[1]{\href{http://ascl.net/#1}{\nolinkurl{http://ascl.net/#1}}}
\providecommand{\doarXiv}[1]{\href{https://arxiv.org/abs/#1}{\nolinkurl{https://arxiv.org/abs/#1}}}

\bibitem[{Bale {et~al.}(2013)Bale, Pulupa, Salem, Chen, \& Quataert}]{Bale2013}
Bale, S.~D., Pulupa, M., Salem, C., Chen, C. H.~K., \& Quataert, E. 2013,
  \apjl, 769, 2, \dodoi{10.1088/2041-8205/769/2/L22}

\bibitem[{{Ber{\v{c}}i{\v{c}}} {et~al.}(2019){Ber{\v{c}}i{\v{c}}},
  {Maksimovi{\'c}}, {}, {Land i}, \& {Matteini}}]{Bercic2019}
{Ber{\v{c}}i{\v{c}}}, L., {Maksimovi{\'c}}, {}, M., {Land i}, S., \&
  {Matteini}, L. 2019, \mnras, 486, 3404, \dodoi{10.1093/mnras/stz1007}

\bibitem[{{Breneman} {et~al.}(2010){Breneman}, {Cattell}, {Schreiner},
  {Kersten}, {Wilson}, {Kellogg}, {Goetz}, \& {Jian}}]{Breneman2010}
{Breneman}, A., {Cattell}, C., {Schreiner}, S., {et~al.} 2010, \jgr, 115,
  A08104, \dodoi{10.1029/2009JA014920}

\bibitem[{Crooker {et~al.}(2003)Crooker, Larson, Kahler, Lamassa, \&
  Spence}]{Crooker2003}
Crooker, N.~U., Larson, D.~E., Kahler, S.~W., Lamassa, S.~M., \& Spence, H.~E.
  2003, \grl, 30, 1619, \dodoi{10.1029/2003GL017036}

\bibitem[{Gary(1985)}]{Gary1985}
Gary, S.~P. 1985, \jgr, 90, 10815, \dodoi{10.1029/JA090iA11p10815}

\bibitem[{{Gary} \& {Feldman}(1977)}]{Gary1977}
{Gary}, S.~P., \& {Feldman}, W.~C. 1977, \jgr, 82, 1087,
  \dodoi{10.1029/JA082i007p01087}

\bibitem[{Gary {et~al.}(1999b)Gary, Neagu, Skoug, \& Goldstein}]{Gary1999b}
Gary, S.~P., Neagu, E., Skoug, R.~M., \& Goldstein, B.~E. 1999b, \jgr, 104,
  19843, \dodoi{10.1029/1999JA900244}

\bibitem[{{Gary} \& {Saito}(2007)}]{Gary2007}
{Gary}, S.~P., \& {Saito}, S. 2007, \grl, 34, L14111,
  \dodoi{10.1029/2007GL030039}

\bibitem[{{Gary} {et~al.}(1999a){Gary}, {Skoug}, \& {Daughton}}]{Gary1999a}
{Gary}, S.~P., {Skoug}, R.~M., \& {Daughton}, W. 1999a, PhPl, 6, 2607,
  \dodoi{10.1063/1.873532}

\bibitem[{{Gurgiolo} {et~al.}(2012){Gurgiolo}, {Goldstein}, {Vi{\~n}as}, \&
  {Fazakerley}}]{Gurgiolo2012}
{Gurgiolo}, C., {Goldstein}, M.~L., {Vi{\~n}as}, A.~F., \& {Fazakerley}, A.~N.
  2012, AnGeo, 30, 163, \dodoi{10.5194/angeo-30-163-2012}

\bibitem[{Horaites {et~al.}(2018)Horaites, Astfalk, Boldyrev, \&
  Jenko}]{Horaites2018}
Horaites, K., Astfalk, P., Boldyrev, S., \& Jenko, F. 2018, \mnras, 480, 1499,
  \dodoi{10.1093/mnras/sty1808}

\bibitem[{{Lacombe} {et~al.}(2014){Lacombe}, {Alexandrova}, {Matteini},
  {Santol{\'\i}k}, {Cornilleau-Wehrlin}, {Mangeney}, {de Conchy}, \&
  {Maksimovic}}]{Lacombe2014}
{Lacombe}, C., {Alexandrova}, O., {Matteini}, L., {et~al.} 2014, \apj, 796, 5,
  \dodoi{10.1088/0004-637X/796/1/5}

\bibitem[{{Landi} {et~al.}(2014){Landi}, {Matteini}, \&
  {Pantellini}}]{Landi2014}
{Landi}, S., {Matteini}, L., \& {Pantellini}, F. 2014, \apj, 790, L12,
  \dodoi{10.1088/2041-8205/790/1/L12}

\bibitem[{Lee {et~al.}(2019)Lee, Lee, \& Yoon}]{Lee2019}
Lee, S.-Y., Lee, E., \& Yoon, P.~H. 2019, \apj, 876, 117,
  \dodoi{10.3847/1538-4357/ab12db}

\bibitem[{L{\'{o}}pez \& Yoon(2018)}]{Lopez2018}
L{\'{o}}pez, R.~A., \& Yoon, P.~H. 2018, \jgr, 123, 8924,
  \dodoi{10.1029/2018JA025934}

\bibitem[{Maksimovic {et~al.}(2005)Maksimovic, Zouganelis, Chaufray, Issautier,
  Scime, Littleton, Marsch, McComas, Salem, Lin, \& Elliott}]{Maksimovic2005}
Maksimovic, M., Zouganelis, I., Chaufray, J.~Y., {et~al.} 2005, \jgr, 110, 1,
  \dodoi{10.1029/2005JA011119}

\bibitem[{Markidis {et~al.}(2010)Markidis, Lapenta, \&
  Rizwan-uddin}]{Markidis2010}
Markidis, S., Lapenta, G., \& Rizwan-uddin. 2010, Math. Comput. Simul., 80,
  1509, \dodoi{10.1016/j.matcom.2009.08.038}

\bibitem[{{Pagel} {et~al.}(2007){Pagel}, {Gary}, {de Koning}, {Skoug}, \&
  {Steinberg}}]{Pagel2007}
{Pagel}, C., {Gary}, S.~P., {de Koning}, C.~A., {Skoug}, R.~M., \& {Steinberg},
  J.~T. 2007, \jgr, 112, A04103, \dodoi{10.1029/2006JA011967}

\bibitem[{{Pavan} {et~al.}(2013){Pavan}, {Vi{\~n}as}, {Yoon}, {Ziebell}, \&
  {Gaelzer}}]{Pavan2013}
{Pavan}, J., {Vi{\~n}as}, A.~F., {Yoon}, P.~H., {Ziebell}, L.~F., \& {Gaelzer},
  R. 2013, \apj, 769, L30, \dodoi{10.1088/2041-8205/769/2/L30}

\bibitem[{Pilipp {et~al.}(1987)Pilipp, Miggenrieder, Mühlhäuser, Rosenbauer,
  Schwenn, \& Neubauer}]{Pilipp1987}
Pilipp, W.~G., Miggenrieder, H., Mühlhäuser, K.~H., {et~al.} 1987, \jgr, 92,
  1103, \dodoi{10.1029/JA092iA02p01103}

\bibitem[{{Saeed} {et~al.}(2017{\natexlab{a}}){Saeed}, {Sarfraz}, {Yoon},
  {Lazar}, \& {Qureshi}}]{Saeed2017a}
{Saeed}, S., {Sarfraz}, M., {Yoon}, P.~H., {Lazar}, M., \& {Qureshi}, M.~N.~S.
  2017{\natexlab{a}}, \mnras, 465, 1672, \dodoi{10.1093/mnras/stw2900}

\bibitem[{{Saeed} {et~al.}(2017{\natexlab{b}}){Saeed}, {Yoon}, {Sarfraz}, \&
  {Qureshi}}]{Saeed2017b}
{Saeed}, S., {Yoon}, P.~H., {Sarfraz}, M., \& {Qureshi}, M.~N.~S.
  2017{\natexlab{b}}, \mnras, 466, 4928, \dodoi{10.1093/mnras/stx049}

\bibitem[{Saito \& Gary(2007a)}]{Saito2007a}
Saito, S., \& Gary, S.~P. 2007a, \grl, 34, L01102, \dodoi{10.1029/2006GL028173}

\bibitem[{Saito \& Gary(2007b)}]{Saito2007b}
---. 2007b, \jgr, 112, A06116, \dodoi{10.1029/2006JA012216}

\bibitem[{{Scime} {et~al.}(2001){Scime}, {Littleton}, {Gary}, {Skoug}, \&
  {Lin}}]{Scime2001}
{Scime}, E.~E., {Littleton}, J.~E., {Gary}, S.~P., {Skoug}, R., \& {Lin}, N.
  2001, \grl, 28, 2169, \dodoi{10.1029/2001GL012925}

\bibitem[{Seough {et~al.}(2015)Seough, Nariyuki, Yoon, \& Saito}]{Seough2015a}
Seough, J., Nariyuki, Y., Yoon, P.~H., \& Saito, S. 2015, \apjl, 811, L7,
  \dodoi{10.1088/2041-8205/811/1/L7}

\bibitem[{Shaaban {et~al.}(2018a)Shaaban, Lazar, \& Poedts}]{Shaaban2018a}
Shaaban, S.~M., Lazar, M., \& Poedts, S. 2018a, \mnras, 480, 310,
  \dodoi{10.1093/mnras/sty1567}

\bibitem[{Shaaban {et~al.}(2018b)Shaaban, Lazar, Yoon, \&
  Poedts}]{Shaaban2018b}
Shaaban, S.~M., Lazar, M., Yoon, P.~H., \& Poedts, S. 2018b, PhPl, 25, 082105,
  \dodoi{10.1063/1.5042481}

\bibitem[{Shaaban {et~al.}(2019)Shaaban, Lazar, Yoon, Poedts, \&
  L{\'{o}}pez}]{Shaaban2019}
Shaaban, S.~M., Lazar, M., Yoon, P.~H., Poedts, S., \& L{\'{o}}pez, R.~A. 2019,
  \mnras, 486, 4498, \dodoi{10.1093/mnras/stz830}

\bibitem[{Spitzer \& H\"arm(1953)}]{Spitzer1953}
Spitzer, L., \& H\"arm, R. 1953, Phys. Rev., 89, 977,
  \dodoi{10.1103/PhysRev.89.977}

\bibitem[{{Stansby} {et~al.}(2016){Stansby}, {Horbury}, {Chen}, \&
  {Matteini}}]{Stansby2016}
{Stansby}, D., {Horbury}, T.~S., {Chen}, C.~H.~K., \& {Matteini}, L. 2016,
  \apjl, 829, L16, \dodoi{10.3847/2041-8205/829/1/L16}

\bibitem[{{Tong} {et~al.}(2018){Tong}, {Bale}, {Salem}, \& {Pulupa}}]{Tong2018}
{Tong}, Y., {Bale}, S.~D., {Salem}, C., \& {Pulupa}, M. 2018, arXiv e-prints,
  arXiv:1801.07694.
\newblock \doarXiv{1801.07694}

\bibitem[{Tong {et~al.}(2019{\natexlab{a}})Tong, Vasko, Artemyev, Bale, \&
  Mozer}]{Tong2019b}
Tong, Y., Vasko, I.~Y., Artemyev, A.~V., Bale, S.~D., \& Mozer, F.~S.
  2019{\natexlab{a}}, \apj, 878, 41, \dodoi{10.3847/1538-4357/ab1f05}

\bibitem[{Tong {et~al.}(2019{\natexlab{b}})Tong, Vasko, Pulupa, Mozer, Bale,
  Artemyev, \& Krasnoselskikh}]{Tong2019a}
Tong, Y., Vasko, I.~Y., Pulupa, M., {et~al.} 2019{\natexlab{b}}, \apjl, 870,
  L6, \dodoi{10.3847/2041-8213/aaf734}

\bibitem[{Vasko {et~al.}(2019)Vasko, Krasnoselskikh, Tong, Bale, Bonnell, \&
  Mozer}]{Vasko2019}
Vasko, I.~Y., Krasnoselskikh, V., Tong, Y., {et~al.} 2019, \apjl, 871, L29,
  \dodoi{10.3847/2041-8213/ab01bd}

\bibitem[{{Verscharen} {et~al.}(2019{\natexlab{a}}){Verscharen}, {Chandran},
  {Jeong}, {Salem}, {Pulupa}, \& {Bale}}]{Verscharen2019b}
{Verscharen}, D., {Chandran}, B. D.~G., {Jeong}, S.-Y., {et~al.}
  2019{\natexlab{a}}, arXiv e-prints.
\newblock \doarXiv{1906.02832}

\bibitem[{{Verscharen} {et~al.}(2019{\natexlab{b}}){Verscharen}, {Klein}, \&
  {Maruca}}]{Verscharen2019}
{Verscharen}, D., {Klein}, K.~G., \& {Maruca}, B.~A. 2019{\natexlab{b}}, arXiv
  e-prints, arXiv:1902.03448.
\newblock \url{https://arxiv.org/abs/1902.03448}

\bibitem[{{Vocks} \& {Mann}(2003)}]{Vocks2003}
{Vocks}, C., \& {Mann}, G. 2003, \apj, 593, 1134, \dodoi{10.1086/376682}

\bibitem[{{Vocks} {et~al.}(2005){Vocks}, {Salem}, {Lin}, \& {Mann}}]{Vocks2005}
{Vocks}, C., {Salem}, C., {Lin}, R.~P., \& {Mann}, G. 2005, \apj, 627, 540,
  \dodoi{10.1086/430119}

\bibitem[{{Wilson} {et~al.}(2013){Wilson}, {Koval}, {Szabo}, {Breneman},
  {Cattell}, {Goetz}, {Kellogg}, {Kersten}, {Kasper}, \& {Maruca}}]{Wilson2013}
{Wilson}, L.~B., {Koval}, A., {Szabo}, A., {et~al.} 2013, \jgr, 118, 5,
  \dodoi{10.1029/2012JA018167}

\end{thebibliography}

\end{document}